\documentstyle[nato,epsfig]{crckapb}

\begin{opening}
\title{Specific heat of Y$_{\lowercase{x}}$L\lowercase{u}$_
{\lowercase{1-x}}$N\lowercase{i}$_2$B$_2$C in the mixed state}
\author{D.\ Lipp, \ M.\  Schneider, A.\ Gladun}
\institute{Institut f\"ur 
Tieftemperaturphysik, Technische Universit\"at Dresden,
\protect{
D-01062 Dresden, Germany}}
\author{S.-L.\ Drechsler,}
\institute{Theor.\ Festk\"orperphysik, 
Inst.\ f.\ Festk\"orper- u.\ Werkstofforschung, 
D-01171 Dresden, Postfach 270116, Germany}
\author{ J.\ Freudenberger, G.\ Fuchs, K.\ Nenkov, K.-H.\ M\"uller}
\institute{Inst.\ f.\ metall. Werkstoffe, Inst.\ f.\
Festk\"orper- u.\ Werkstofforschung, D-01171 Dresden, Postfach 270116, Germany}
\author{T.\ Cichorek, P.\ Gegenwart}
\institute{Max-Planck-Institut f\"ur Chemische 
Physik fester Stoffe Dresden,\\
D-01187 Dresden, Germany}
\end{opening}

\begin{document}
\noindent
{\footnotesize
The temperature and magnetic field dependences of the electronic 
specific heat $c_p$ in the superconducting mixed state as well as  
the upper critical field $H_{c2}(T)$   have been measured 
 on a 
series of polycrystalline Y$_x$Lu$_{1-
x}$Ni$_2$B$_2$C  samples 
  with the aim to study the influence of 
disorder.   
The electronic 
specific heat contribution $c_{esm}=\gamma(H)\cdot T$ in 
the mixed (vortex) state exhibits significant deviations 
from the usual  
$\gamma (H) \propto H$-law for conventional 
superconductors for all compositions  
resulting in a disorder  dependent 
negative curvature of $\gamma \propto H^{1-\beta(x)}$.
 For the  samples 
 with substitutional disorder at the rare earth 
 site $\beta$ and  the positive curvature exponent 
 $\alpha$ of $H_{c2}(T)$ are  reduced but
  $\alpha, \beta > 0$ holds for all $x$, as distinct from  
 the case of much stronger disorder due to
isoelectronic partial Pt substitutions at the Ni site 
 where $\alpha$ and $ \beta$ rapidly  vanish (Nohara {\it et al.} (1999) 
 {\it J.\ Phys.\ Soc.\ Jpn.\ }{\bf 68} 1078).
   $\alpha$  exhibits 
 a clear minimum near $x= 0.5$. 
The magnitudes of $H_{c2}(T)$ and $T_c$, the curvatures of $H_{c2}(T)$ and  
 $\gamma(H)$,  as well as the 
Sommerfeld constant $\gamma_N$ in the normal state
 are reduced in a similar fashion for intermediate composition.}
 
\noindent

\section{Introduction}

The recent discovery of the  rare earth transition metal 
borocarbide (nitride)  
 family
(RC)$_n$T$_2$B$_2$C(N), R=Y, Lu, Sc, Th, La; T=Ni, Pd, Pt;
$n$=1 to 3  
\cite{Nagarajan,Cava2} which contains  superconductors with 
relatively 
high transition temperatures 
$T_c$ up to 23K has stimulated numerous 
studies of their thermodynamic and 
transport properties 
in the superconducting as well as in the normal state. 
At first glance most of  those results support  
a classification 
of these materials as typical intermetallic 
phonon mediated 
superconductors with a moderately strong 
coupling strength. However, clean
RNi$_2$B$_2$C samples
 exhibit also
some features unexpected for ordinary
superconductors both in the superconducting and 
in the normal state. Among several 
peculiarities \cite{drechsler98}
 we emphasize the unusual shape and the
 strong disorder dependence of 
 the upper critical magnetic 
field $H_{c2}(T)$ and the $T^3$ dependence 
of the electronic specific heat $c_{es}(T)$ in 
the superconducting state compared with
exponential behaviour  for 
ordinary $s$-wave superconductors. According to 
Nohara {\em et al.\/} \cite{Nohara1} 
isoelectronic transition metal substitution affects   strongly
the  field dependence of the 
electronic specific heat contribution $c_{esm}=\gamma(H)\cdot 
T$  in the mixed state. While for an 
Y(Ni$_{0.8}$Pt$_{0.2})_2$B$_2$C 
single crystal $\gamma (H)\propto H$, a marked 
sublinear law was observed for pure
YNi$_2$B$_2$C single crystal and polycrystalline 
LuNi$_2$B$_2$C   \cite{Nohara2}
\begin{equation}
\gamma(H)/\gamma_N \propto \sqrt{H/H_{c2}(0)},
\end{equation}
where $\gamma_N$ denotes the 
 Sommerfeld constant in the normal state.
 Thus already small 
substitutions  in the  TB network (which is 
commonly believed to be
the crucial locus for
the superconductivity) 
affect significantly the superconducting 
properties, pressumably due to the predominant role of the 
T-derived $d$ states in the mechanism of superconductivity. 
At the same time  by isoelectronic substitutions  in the RC charge reservoir 
 much weaker disorder could be studied (see Fuchs {\it et al.}, Drechsler
 {\it et al.}, these Proc.). 
The remarkable $\gamma (H) \propto \sqrt{H}$-law  for YNi$_2$B$_2$C 
 and LuNi$_2$B$_2$C
 was regarded at first as evidence for $d$-wave pairing 
\cite{Nohara2}-\cite{Maki}. In this context it is interesting to study also 
the closely related   Y$_x$Lu$_{1-x}$Ni$_2$B$_2$C system as 
an appropriate weak disorder case. Thus changing the composition $x$,    
  more detailed information should be gained about the influence of  weak   
disorder  
upon the field dependence 
of the specific heat $c_p(T,H)$,  
 the shape and the magnitude of $H_{c2}(T)$,
 as well as about the nature of the pairing state.

\section{Experimental Details}
 Polycrystalline Y$_x$Lu$_{1-x}$Ni$_2$B$_2$C 
  samples were prepared by a standard arc melting
technique. Powders of the elements were weighted in the stoichiometric
compositions with a surplus of 10 wt.\% boron to compensate the high
losses of boron caused by the arc melting.  The powder was pressed to   
pellets which were melted under argon gas on a water-cooled copper plate
in an arc furnace. To get homogeneous samples, they were turned over and
melted again four times. After the melting procedure the solidified
samples were homogenised at $1100^{\circ}\rm C$\ for ten days.                              
 The specific heat was measured in the 
 range 4.2K $\le T \le $ 20K
and for magnetic fields $H \leq$  8T using a quasi 
adiabatic step heating technique.
The heating pulses were generated by a strain 
gauge heater and $T$ was measured
with an Au-Ge thin film resistor.

\begin{figure}[hbt]
\vspace{-0.6cm}
\begin{center}
\epsfig{file=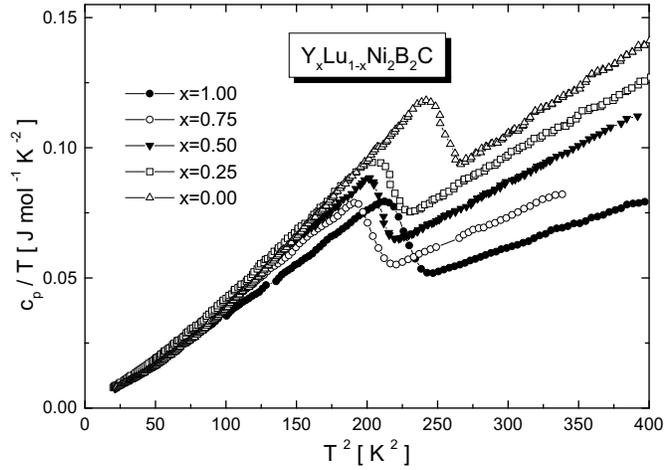,width=0.75\textwidth}
\caption{Zero magnetic field specific heat  
$c_p(T)/T$ vs. $T^2$ of a Y$_x$Lu$_{1-
x}$Ni$_2$B$_2$C series. 
}
\label{fig1}
\end{center}
\end{figure}
\vspace{-0.5cm}
\begin{figure}[hbt]
\begin{center}
\epsfig{file=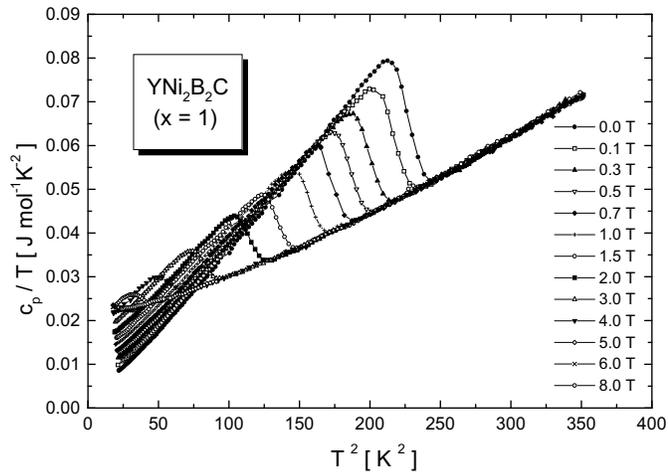,width=0.75\textwidth}
\caption{Specific heat $c_p(T,H)/T$ vs.\ $T^2$ of 
YNi$_2$B$_2$C for various magnetic 
fields.}
\label{fig2}
\end{center}
\vspace{-0.5cm}
\end{figure}
\section{Results and Discussion}

The specific heat $c_p/T$ vs. $T^2$  at $H = 0$ 
of a Y$_x$Lu$_{1-x}$Ni$_2$B$_2$C series  
is shown in 
Fig.\ \ref{fig1}. The  corresponding 
curves for $H \leq$  8T of the 
pure Y sample ($x=1$)
are depicted in 
Fig.\ \ref{fig2}.  
Applying $H= 8$T
 resulted in a 
shift of $T_c$ below the lowest 
measured temperature of 4.2K.
Measurements in this field were used to analyze 
the    
normal state specific heat $c_p=\gamma _N T + 
\beta_D T^3$, where $\gamma _N$ is the Sommerfeld 
constant  and $\beta_D T^3$ is the Debye contribution. 
The  $\gamma_N$ was 
determined by extrapo-

 \begin{figure}[hbt].
\begin{center}
\epsfig{file=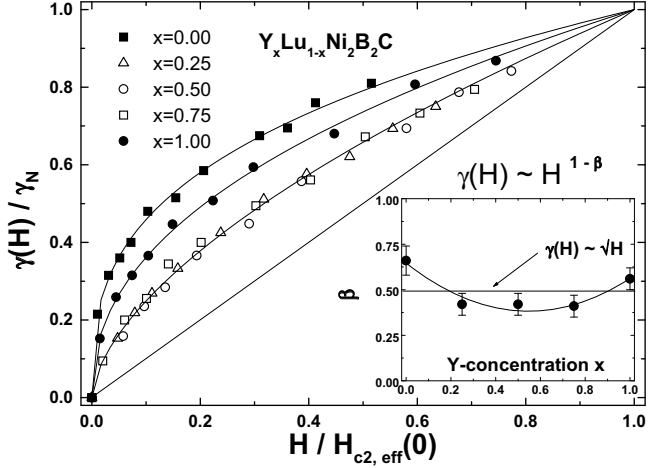,width=0.75\textwidth}
\caption{Magnetic field dependence of the   
specific heat contribution 
$\gamma (H)$ of the vortex core electrons in the 
mixed state ($H \le H_{c2}$) in units of the effective upper critical 
field $H_{c2,eff}(0)$
shown in Fig.\ 4. The 
straight reference line corresponds to the usual $s$-wave dirty limit 
behaviour. 
The inset shows the curvature parameter $\beta$ 
of $\gamma (H)$ according to
$\gamma (H)/\gamma _N \propto (H/H_{c2, eff}(0))^{1-\beta}$. Noteworthy, 
the case $\beta$ = 0.31 to 0.35 and $H/H_{c2, eff}(0)\leq 0.75$
  is not very distinct  from the dirty (unitary) $d$-wave limit 
 where $\gamma(H)\propto H\ln H$ according to Ref.\ [17] (not shown here).}
\label{fig4}
\end{center}
\end{figure}
 \vspace{-0.0cm}
\noindent 
lating the $c_p/T$ vs 
$T^2$ curves to $T = 0$ in the normal state. 
This way we obtained $\gamma _N$= 
20.4 ($x=0$), 19 ($x=0.25$), 
18.3 ($x=0.5$), 18 ($x=0.75$) and 20.2mJ/molK$^2$
 ($x=1$) for our 
Y$_x$Lu$_{1-x}$Ni$_2$B$_2$C series in good 
agreement with  the data reported by several groups
\cite{Manalo}-\cite{Carter}. 
To      determine
  $\gamma (H)$, the $c_p/T$ vs.\ $T^2$ 
curves have been extrapolated linearly
to 
$T=0$ from the data in the range 4.2K $\le T \le 7 $K or up to 
the onset of the transition to the normal state. 
The results are shown in Fig.\ \ref{fig4}. 
 Apparently, for all Y$_x$Lu$_{1-
x}$Ni$_2$B$_2$C samples $\gamma (H)$ is a
 sublinear function on $H$. Generalizing Eq.\ (1), the data were analyzed
 by the expression 
\begin{equation}
\gamma (H)/\gamma_N = (H/H_{c2,eff}(0,x))^{1-\beta(x)},
\end{equation} where the fitting 
parameter $\beta$ measures  the negative 
curvature of $\gamma (H)$ while $H_{c2,eff}(0)\approx $ 
(0.8 to 0.9)$H_{c2}(0)$ yields 
a lower
 bound for the true $H_{c2}(0)$ (see Fig.\ 4). We obtained
$\beta =$ 0.66, 0.42, 0.42, 0.41 and 0.56 from 
$x=$ 0 to $x=$ 1, 
respectively. The behaviour of $\beta (x)$  
 is shown in the 
inset of Fig.\ \ref{fig4}. The curvature of 
$\gamma (H)$ becomes maximal
 for the bordering cases $x=$ 0; 1 
and markedly smaller in between. Noteworthy,  
our curvatures for LuNi$_2$B$_2$C 
and YNi$_2$B$_2$C do  slightly exceed
 those  derived from Eq.\ (1) as  reported in Refs.\ 
 \cite{Nohara1,Nohara2}. 
Anyhow, $\gamma (H)$ 
shows for
all  Y$_x$Lu$_{1-x}$Ni$_2$B$_2$C samples  a clear
$H$ sublinear dependence. A $\gamma (H) \approx \sqrt{H}$-law
as well as the $c_{es}(T) \propto T^3$ 
dependence raise the question 
whether an 
unconventional pairing mechanism is responsible for this
peculiarity. According to Ref.\ \cite{Volovik}
$\gamma (H) \propto \sqrt{H}$ 
 is a signature 
for a nodal order parameter with $d$-wave pairing (for
 which
$\beta =0.57$ has been predicted in Ref.\ \cite{ichioka99}) while
 $\gamma (H) 
\propto \ H$ points to  nodeless  
$s$-wave pairing in the dirty limit since 
there are  conventional 
superconductors which exhibit  
also deviations 
from the  
$\gamma (H) \propto H$-law in the clean limit,  e.g.\  
V$_3$Si \cite{Ramirez}, NbSe$_2$\cite{Nohara1} 
($\beta=0.33$),
 and CeRu$_2$\cite{ichioka99}. 
Hence, the 
 sublinear $\gamma 
(H)$ dependences
 observed here are
  not inevitably 
 related to $d$-wave pairing \cite{remark}.

\begin{figure}[h!]
\vspace{-0.3cm}
\begin{center}
\epsfig{file=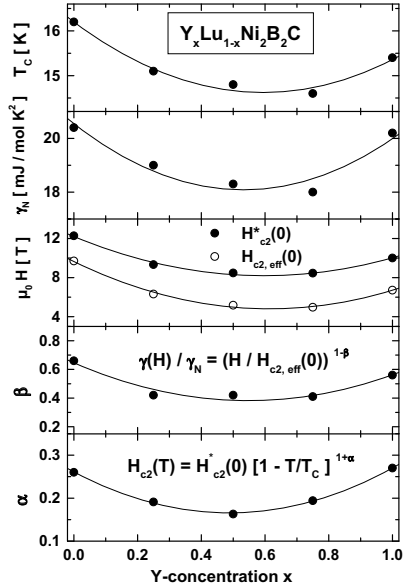,width=0.47\textwidth}
\caption{Composition dependence of the 
transition temperature $T_c$ derived from the 
onset 
of the jump of $c_P(T)$ (upper panel), the 
Sommerfeld constant $\gamma _N$ (second panel), 
the upper (lower) bound for the upper critical
 field (third panel, see Eqs.\ (2,3) and text),
the specific heat curvature exponent  $\beta$ of 
$\gamma(H)$  (fourth panel), and the 
curvature exponent $\alpha$ of the upper critical field 
$H_{c2}$  (lower 
panel).}
\label{fig6}
\end{center}
\vspace{-0.5cm}
\end{figure}

Since the lattice constants vary 
linearly between $x=0$ and $x=1$, one might
expect a maximum of $T_c$ around $x=0.6$  
according to an
 ``universal'' curve $T_c = T_c(d_{Ni-Ni})$ 
proposed by Lai {\em et al.\/} \cite {Lai},
where $d_{Ni-Ni}$ is the Ni-Ni distance. 
However, in accord with Refs.\ [8,9] 
 a dip near 
$x=0.7$ is observed ($T_c \approx 14.6$K at $x=0.75$). A 
similar behaviour is found
for $H_{c2}(0)$ and also for the curvature exponent $\alpha$ 
(see Fig. \ref{fig6}):
\begin{equation}
H_{c2}(T)=H^*_{c2}(1-T/T_c)^{1+\alpha}, \quad \mbox{valid for} \quad 
0.3 \leq T/T_c ,
\end{equation}
(distinct from a statement of a constant curvature \cite{Manalo}).
$H^*_{c2}\approx 1.1 \div 1.2 H_{c2}(0)$ gives an upper bound for the true
value $H_{c2}(0)$ at $T=0$. Ongoing from pure  samples in the clean limit
to the mixed specimens which remain still in the quasi-   
clean limit,
 all these quantities are somewhat 
reduced due to weak disorder 
 at 
\vspace{0.0cm} 
\noindent
 the R-sites as well as  due to  
a slight reduction of the electron-phonon coupling constant $\lambda$ 
  at intermediate 
$x$ deduced from the similar 
behavior of  $\gamma_N$
\cite{drechsler99a,freudenberger99} (see Fuchs 
{\it et al.,} these Proc.).

To summarize, we have shown that the $\gamma 
(H)$ curves of the pure specimens ($x=$ 0; 1)
of Y$_x$Lu$_{1-x}$Ni$_2$B$_2$C exhibit the 
highest negative curvatures. Weak disorder effects,
caused by isoelectronic substitutions of Y by 
Lu, yield a decrease of the curvature  without
reaching the linear limit,  which would be the hallmark
for the dirty limit.
Similar moderate suppressions which are typical for the 
quasi-clean limit
have been found for the upper critical field,
 its curvature,  and  $T_c$.  

\vspace{-0.3cm}

\begin{acknowledgements}
\vspace{-0.3cm}
This work was supported by the SFB 463
and the DFG.
We  acknowledge discussions with S.\ Shulga,
H.\ Rosner, H.\ Michor, M.\ Nohara, K.\ Maki, and H.\ Takagi.
\end{acknowledgements}


\begin{thebibliography}{000}
\bibitem{remark}
However, see also the note in the caption of Fig.\ 3.
\end{thebibliography}
\end{document}